# Implication of Compensator Field and Local Scale Invariance in the Standard Model

Hitoshi NISHINO[1)] and Subhash RAJPOOT[2)]

*Department of Physics and Astronomy*
*California State University*
*1250 Bellflower Boulevard*
*Long Beach, CA 90840*

**Abstract**

We introduce Weyl's scale symmetry into the standard model (SM) as a local symmetry. This necessarily introduces gravitational interactions in addition to the local scale invariance group $\widetilde{U}(1)$ and the SM groups $SU(3) \times SU(2) \times U(1)$. The only other new ingredients are a new scalar field $\sigma$ and the gauge field for $\widetilde{U}(1)$ we call the Weylon. A noteworthy feature is that the system admits the Stückelberg-type compensator. The $\sigma$ couples to the scalar curvature as $(-\zeta/2)\,\sigma^2 R$, and is in turn related to a Stückelberg-type compensator $\varphi$ by $\sigma \equiv M_{\rm P} e^{-\varphi/M_{\rm P}}$ with the Planck mass $M_{\rm P}$. The particular gauge $\varphi = 0$ in the Stückelberg formalism corresponds to $\sigma = M_{\rm P}$, and the Hilbert action is induced automatically. In this sense, our model presents yet another mechanism for breaking scale invariance at the classical level. We show that our model naturally accommodates the chaotic inflation scenario with no extra field.

[1)] E-Mail: hnishino@csulb.edu

[2)] E-Mail: rajpoot@csulb.edu

## 1. Introduction

We consider Weyl's original idea on local scale invariance [1] in the context of the extension of the standard model (SM). This necessarily requires gravitational interactions with the diffeomorphism group to be treated on par with the other particle interactions. The symmetry of our action is (Diffeomorphisms) $\times SU(3) \times SU(2) \times U(1) \times \widetilde{U}(1)$, where $\widetilde{U}(1)$ is for local scale invariance.

However, scale invariance symmetry is broken symmetry in Nature. In this Letter, we investigate the breaking of local scale invariance [1], *via* the Stückelberg mechanism [2][3][4]. The Stückelberg extension of the SM has been recently considered for Abelian gauge groups [5]. Our work is similar in spirit in the sense that we introduce local scale invariance group $\widetilde{U}(1)$ in addition to the standard groups $SU(3) \times SU(2) \times U(1)$.

However, the difference in our work from [5] is that scale invariance also acts on the space-time metric. It is also different from non-linear realization of scale invariance [6][7]. The most noteworthy feature of our model is the economy in extending the SM with only a very limited number of additional fields. The new fields added to the particle spectrum of the SM are the graviton $e_\mu{}^m$, a vector boson $S_\mu$ we call the Weylon, and a real singlet scalar field $\sigma$, where $\sigma$ will be eventually absorbed into the longitudinal component of $S_\mu$.

Our total field content consists of the usual vierbein $e_\mu{}^m$, the quarks and leptons $\Psi^{gf}$, $\Psi_i^{gf}$,[3)] the Higgs doublet $\Phi$, the single real scalar $\sigma$, and the gauge field $S_\mu$ for $\widetilde{U}(1)$ we call the Weylon, and the usual SM gauge fields $A_\mu$, $W_\mu$ and $B_\mu$ for the gauge groups $SU(3)$, $SU(2)$ and $U(1)$, respectively. Under $\widetilde{U}(1)$ these fields transform as [8]

$$\begin{aligned}
& e_\mu{}^m \longrightarrow e^{+\Lambda} e_\mu{}^m \ , \quad g_{\mu\nu} \longrightarrow e^{+2\Lambda} g_{\mu\nu} \ , \quad e \equiv \sqrt{-g} \longrightarrow e^{+4\Lambda} \sqrt{-g} \ , \\
& \Psi^{gf} \longrightarrow e^{-3\Lambda/2}\, \Psi^{gf} \ , \quad \Psi_i^{gf} \longrightarrow e^{-3\Lambda/2}\, \Psi_i^{gf} \ , \quad \Phi \longrightarrow e^{-\Lambda}\, \Phi \ , \quad \sigma \longrightarrow e^{-\Lambda}\, \sigma \ , \\
& S_\mu \to S_\mu - f^{-1} \partial_\mu \Lambda \ , \qquad (1.1)
\end{aligned}$$

with the finite local scale transformation parameter $\Lambda \equiv \Lambda(x)$, while $A_\mu$, $W_\mu$ and $B_\mu$ are invariant.

Before presenting our lagrangian, we stress the basic difference from past works in the

---

[3)] The indices $g = 1, 2, 3$ are for the three generations, $f = q, l$ are for the quarks and leptons, while $i = 1, 2$ are needed for the right-handed fermions.

literature on dilaton[4] or scalar field coupled to scalar curvature, in order to avoid possible confusion. For example in [9], a scalar $\varphi$ is coupled to the scalar curvature like $(1/2)\varphi^2 R$ together with a potential $V(\varphi)$, such that $\langle\varphi\rangle = v$ will yield the Newton's gravitational constant. However, *no* local scale invariance was required in [9]. As other examples, in [10] a dilaton-scalar curvature coupling is considered, while in [11] fairly general couplings of a singlet and Higgs doublet to a scalar curvature are considered, but there was *no* introduction of local scale invariance, with *no* gauge field, as opposed to our system with $S_\mu$. In ref. [12], local scale invariance is considered, even *without* its gauge field, because the usual kinetic term for a Dirac field possesses *local* scale invariance *without* the Weylon. In our present paper, we introduce *local* scale invariance (1.1) with its gauge field $S_\mu$, coupled also to the SM system, which is clearly distinct from any past work on global scale invariance.

Our total action invariant under (Diffeomorphisms) $\times SU(3)\times SU(2)\times U(1)\times \widetilde{U}(1)$ is [8]

$$\begin{aligned} I = \int d^4x\, e \bigg[ & -\tfrac{1}{2}(\beta\Phi^\dagger\Phi + \zeta\sigma^2)\widetilde{R} - \tfrac{1}{4}g^{\mu\rho}g^{\nu\sigma}\Big\{ \mathrm{Tr}\,(W_{\mu\nu}W_{\rho\sigma}) + B_{\mu\nu}B_{\rho\sigma} + U_{\mu\nu}U_{\rho\sigma}\Big\} \\ & + \sum_{\substack{\mathrm{f}=q,l \\ \mathrm{g}=1,2,3}} \Big( \overline{\Psi}{}^{\mathrm{gf}}_{\mathrm{L}}\,\gamma^\mu D_\mu \Psi^{\mathrm{gf}}_{\mathrm{L}} + \sum_{i=1,2} \overline{\Psi}{}^{\mathrm{gf}}_{i\mathrm{R}}\,\gamma^\mu D_\mu \Psi^{\mathrm{gf}}_{i\mathrm{R}} \Big) + \sum_{\substack{\mathrm{f}=q,l \\ \mathrm{g,g'}=1,2,3 \\ i=1,2}} \Big( \mathbf{Y}^{\mathrm{f}}_{\mathrm{gg'}} \overline{\Psi}{}^{\mathrm{gf}}_{\mathrm{L}}\Phi\Psi^{\mathrm{g'f}}_{i\mathrm{R}} + \mathbf{Y}'^{\mathrm{f}}_{\mathrm{gg'}}\overline{\Psi}{}^{\mathrm{gf}}_{\mathrm{L}}\widetilde{\Phi}\,\Psi^{\mathrm{g'f}}_{i\mathrm{R}} \Big) + \mathrm{h.c.} \\ & + g^{\mu\nu}(D_\mu\Phi)(D_\nu\Phi^\dagger) + \tfrac{1}{2}g^{\mu\nu}(D_\mu\sigma)(D_\nu\sigma) - \lambda(\Phi^\dagger\Phi)^2 + \mu(\Phi^\dagger\Phi)\sigma^2 - \xi\sigma^4 \bigg] \ , \end{aligned} \tag{1.2}$$

where $\gamma^\mu \equiv \gamma^m e_m{}^\mu$, and any $SU(3)$ color-related terms and indices are suppressed. The field strengths $W_{\mu\nu}$ and $B_{\mu\nu}$ are respectively those of $W_\mu$ and $B_\mu$, while $U_{\mu\nu} \equiv \partial_\mu S_\nu - \partial_\nu S_\mu$. These field strengths are all invariant under $\widetilde{U}(1)$. The scale-invariant scalar curvature $\widetilde{R} \equiv g^{\mu\nu}\widetilde{R}_{\mu\nu}$ and the Ricci tensor $\widetilde{R}_{\mu\nu} \equiv \widetilde{R}_{\mu\rho\nu}{}^\rho$ are defined in terms of the scale-invariant Riemann tensor $\widetilde{R}_{\mu\nu\rho}{}^\sigma \equiv \partial_\mu\widetilde{\Gamma}_{\nu\rho}{}^\sigma - \partial_\nu\widetilde{\Gamma}_{\mu\rho}{}^\sigma - \widetilde{\Gamma}_{\mu\rho}{}^\tau\widetilde{\Gamma}_{\nu\tau}{}^\sigma + \widetilde{\Gamma}_{\nu\rho}{}^\tau\widetilde{\Gamma}_{\mu\tau}{}^\sigma$, where the scale-invariant affinity $\widetilde{\Gamma}$ is defined by $\widetilde{\Gamma}_{\mu\nu}{}^\rho \equiv (1/2)g^{\rho\sigma}(D_\mu g_{\nu\sigma} + D_\nu g_{\mu\sigma} - D_\sigma g_{\mu\nu})$ with $D_\mu g_{\rho\sigma} \equiv \partial_\mu g_{\rho\sigma} + 2fS_\mu g_{\rho\sigma}$. The $\widetilde{\Phi}$ is $\widetilde{\Phi} \equiv i\sigma_2\Phi^\dagger$ and the scale-covariant derivative $D_\mu$ is defined on each field by

$$D_\mu\Psi^{\mathrm{gf}}_L = \Big(\partial_\mu + ig\tau\cdot W_\mu + \tfrac{i}{2}g'\,Y^{\mathrm{gf}}_{\mathrm{L}}\,B_\mu - \tfrac{1}{4}\widetilde{\omega}_\mu{}^{mn}\gamma_{mn} - \tfrac{3}{2}fS_\mu\Big)\Psi^{\mathrm{gf}}_L \ , \tag{1.3a}$$

$$D_\mu\Psi^{\mathrm{gf}}_{iR} = \Big(\partial_\mu + \tfrac{i}{2}g'\,Y^{\mathrm{gf}}_{i\mathrm{R}}\,B_\mu - \tfrac{1}{4}\widetilde{\omega}_\mu{}^{mn}\gamma_{mn} - \tfrac{3}{2}fS_\mu\Big)\Psi^{\mathrm{gf}}_{iR} \ , \tag{1.3b}$$

$$D_\mu\Phi = \Big(\partial_\mu + ig\tau\cdot W_\mu - \tfrac{i}{2}g'B_\mu - fS_\mu\Big)\Phi \ , \qquad D_\mu\sigma = (\partial_\mu - fS_\mu)\,\sigma \ , \tag{1.3c}$$

[4] We can regard our scalar field $\sigma$ as a 'dilaton'. However, the word 'dilaton' is used in the context of global transformation, which we would like to avoid in this paper. We will come back to this in section 2.

where the generators $\tau$ are for $SU(2)$. The scale-invariant Lorentz connection $\widetilde{\omega}_\mu{}^{rs}$ is defined by $\widetilde{\omega}_{mrs} \equiv (1/2)(\widetilde{C}_{mrs} - \widetilde{C}_{msr} + \widetilde{C}_{srm})$, where $\widetilde{C}_{\mu\nu}{}^r \equiv D_\mu e_\nu{}^r - D_\nu e_\mu{}^r$ and $D_\mu e_\nu{}^r \equiv \partial_\mu e_\nu{}^r + f S_\mu e_\nu{}^r$.

For readers who are bothered by the absence of the imaginary unit 'i' in front of the Weylon term in (1.3c) compared with the $U(1)$ coupling by $B_\mu$, we give the following simple justification. The usual $U(1)$ current of a complex scalar $\Phi$ is given by

$$J_\mu = i \left[ \Phi^\dagger \partial_\mu \Phi - (\partial_\mu \Phi^\dagger) \Phi \right] \quad . \tag{1.4}$$

This current $J_\mu$ is hermitian under a complex conjugation, because two terms within the square brackets replace each other with opposite sign, while that sign flip is compensated by the imaginary unit $i \to -i$. In the case of our $\widetilde{U}(1)$, the corresponding current is

$$K_\mu = \Phi^\dagger \partial_\mu \Phi + (\partial_\mu \Phi^\dagger) \Phi \quad . \tag{1.5}$$

Note the relatively positive sign between these two terms, and the absence of the imaginary unit in front. Under a complex conjugation, the two terms replace each other, without any sign flip. This also justifies the absence of the imaginary unit in front. Due to this feature, there is *no* hermiticity problem with the minimal coupling of the Weylon $S_\mu$ to the scalar fields $\Phi$ and $\sigma$ at the lagrangian level.

Note that the potential terms[5] as the last three terms in (1.2) are the most general $\widetilde{U}(1)$- and $SU(2)$-invariant polynomial combinations of $\Phi$ and $\sigma$. Additionally, the terms in the first line with $\Phi$, $\sigma$ and $\widetilde{R}$ are the most general scale invariant combinations. In our previous work on scale invariance in the SM [8], local scale invariance [1] was broken 'by hand' with $\langle \sigma \rangle = \Delta/\sqrt{2}$. In this paper, we provide a simple scheme to achieve the same goal.

## 2. Expressing $\sigma$ in Terms of Compensator $\varphi$

We now show that the original $\sigma$-field is rewritten in terms of a 'dilaton',[6] which plays the role of a compensator for local scale symmetry. The $\sigma$ and the 'dilaton' are related by

$$\sigma = \zeta^{-1/2} M_{\rm P} \, e^{-\kappa\varphi} \quad , \tag{2.1}$$

[5] Let us symbolize these potential terms by $-V(\Phi, \sigma)$.

[6] We use the quotation marks for 'dilaton', because in our system the usual *global* dilaton-shift symmetry $\varphi \to \varphi + const.$ is replaced by the *local* one (2.2).

where $M_{\rm P} \equiv 1/\sqrt{8\pi G_{\rm N}} \simeq 2.44\times 10^{18}\,{\rm GeV}$, and as usual in gravitational theory, $\kappa \equiv 1/M_{\rm P}$ is the natural unit providing the dimension of $({\rm mass})^{-1}$. This choice of $\varphi$ is very natural, because $\sigma$ transforms as $\sigma \rightarrow e^{-\Lambda}\sigma$, while the 'dilaton' $\varphi$ transforms under $\widetilde U(1)$ as

$$\varphi \quad \longrightarrow \quad \varphi + M_{\rm P}\Lambda(x) \quad . \tag{2.2}$$

Rewriting $\sigma$ in terms of $\varphi$ everywhere in the lagrangian (1.2), we get the action

$$\begin{aligned} I = \int d^4x \; e \, \Bigg[ & - \tfrac{1}{2}(\beta\Phi^\dagger\Phi + \zeta M_{\rm P}^2\, e^{-2\kappa\varphi})\,\widetilde R - \tfrac{1}{4}\,{\rm Tr}\left\{(W_{\mu\nu})^2\right\} - \tfrac{1}{4}(B_{\mu\nu})^2 - \tfrac{1}{4}(U_{\mu\nu})^2 \\ & + \sum_{\substack{{\rm f}=q,l \\ {\rm g}=1,2,3}} \Big( \overline\Psi{}^{\rm gf}_{\rm L}\,\gamma^\mu D_\mu \Psi^{\rm gf}_{\rm L} + \sum_{i=1,2} \overline\Psi{}^{\rm gf}_{i{\rm R}}\,\gamma^\mu D_\mu \Psi^{\rm gf}_{i{\rm R}} \Big) + \sum_{\substack{{\rm f}=q,l \\ {\rm g},{\rm g}'=1,2,3 \\ i=1,2}} \Big( {\bf Y}^{\rm f}_{{\rm g}{\rm g}'}\, \overline\Psi{}^{\rm gf}_{\rm L}\Phi\Psi^{{\rm g}'{\rm f}}_{i{\rm R}} + {\bf Y}'^{\rm f}_{{\rm g}{\rm g}'}\, \overline\Psi{}^{\rm gf}_{\rm L}\widetilde\Phi\,\Psi^{{\rm g}'{\rm f}}_{i{\rm R}} \Big) + {\rm h.c.} \\ & + g^{\mu\nu}(D_\mu\Phi^\dagger)(D_\nu\Phi) + \tfrac{1}{2}e^{-2\kappa\varphi}(D_\mu\varphi)^2 - \lambda(\Phi^\dagger\Phi)^2 + \mu M_{\rm P}^2\, e^{-2\kappa\varphi}\,(\Phi^\dagger\Phi) - \xi M_{\rm P}^4\, e^{-4\kappa\varphi} \Bigg] \; , \end{aligned} \tag{2.3}$$

where $D_\mu\varphi \equiv \partial_\mu\varphi + f M_{\rm P} S_\mu$, which is invariant (stronger than covariant) under our $\widetilde U(1)$.

If we redefine $S_\mu$ by $\check S_\mu \equiv S_\mu + f^{-1}\kappa\,\partial_\mu\varphi$, then $D_\mu\varphi = fM_{\rm P}\check S_\mu$, and the kinetic term of $\varphi$ becomes the mass term of $\check S_\mu$:

$$\tfrac{1}{2}e^{-2\kappa\varphi}(D_\mu\varphi)^2 = \tfrac{1}{2}(fM_{\rm P})^2\, e^{-2\kappa\varphi}(\check S_\mu)^2 \quad , \tag{2.4}$$

while the $\check S_\mu$-kinetic term stays form invariant: $-\,(1/4)(U_{\mu\nu})^2 = -(1/4)(\check U_{\mu\nu})^2$. All the covariant derivatives are now

$$\begin{aligned} D_\mu\Psi^{\rm gf}_L = & \left[ \partial_\mu + igW_\mu\cdot\tau + \tfrac{i}{2}g'\,Y^{\rm gf}_{\rm L}\,B_\mu - \tfrac{3}{2}f\check S_\mu + \tfrac{3}{2}\kappa\partial_\mu\varphi - \tfrac{1}{4}\widetilde\omega_\mu{}^{mn}\gamma_{mn} \right] \Psi^{\rm gf}_L \quad , \\ D_\mu\Psi^{\rm gf}_{iR} = & \left[ \partial_\mu + \tfrac{i}{2}g'\,Y^{\rm gf}_{i{\rm R}}\,B_\mu - \tfrac{3}{2}f\check S_\mu + \tfrac{3}{2}\kappa\partial_\mu\varphi - \tfrac{1}{4}\widetilde\omega_\mu{}^{mn}\gamma_{mn} \right] \Psi^{\rm gf}_{i{\rm R}} \quad , \\ D_\mu\Phi = & \left[ \partial_\mu + igW_\mu\cdot\tau - \tfrac{1}{2}g'B_\mu - f\check S_\mu + \kappa\partial_\mu\varphi \right] \Phi \quad . \end{aligned} \tag{2.5}$$

If $\partial_\mu\Lambda = 0$, the $\varphi$-field is essentially the usual dilaton also used in string theory [13]. However, in our system $\varphi$ is not really a dilaton, and it serves as the Stückelberg-type *compensator* [2] under $\widetilde U(1)$, as (2.2) clearly shows.

Since we are dealing with a Stückelberg system [2], there must be a convenient frame where the compensator $\varphi$ vanishes. In fact, we can consider a particular scale transformation with the parameter $\Lambda = -\kappa\varphi$, such that the transformed field of $\varphi$ becomes exactly zero: $\varphi \rightarrow \varphi + M_{\rm P}\Lambda = \varphi - \varphi = 0$, In this case, the Weylon field $\check S_\mu$ is *invariant*, because $\check{S}_\mu = f^{-1}\kappa D_\mu\varphi$ with the manifestly invariant derivative $D_\mu\varphi$. Therefore, under this special transformation $\Lambda = -\kappa\varphi$, the lagrangian (2.3) transforms to the frame, where $\varphi$ in all the exponents is set to *zero*, while the $\varphi$-kinetic term becomes the mass term of $\check{S}_\mu$. Also, $\varphi$-dependent terms in the covariant derivatives in (2.5) disappear. In terms of expressions in our original paper [8], all of these are equivalent to $\varphi = 0 \Longrightarrow \sigma = M_{\rm P}$ and $\Delta \equiv \sqrt{2}M_{\rm P}$, with the $\sigma$-field now eaten up by the Weylon. To be more explicit, our final action is

$$\begin{aligned} I \;=& \int d^4x\, e \bigg[ -\tfrac{1}{2}(\beta\Phi^\dagger\Phi + M_{\rm P}^2)\,\widetilde{R} - \tfrac{1}{4}\,{\rm Tr}\left\{(W_{\mu\nu})^2\right\} - \tfrac{1}{4}(B_{\mu\nu})^2 - \tfrac{1}{4}(\check{U}_{\mu\nu})^2 \\ &+ \sum_{\substack{{\rm f}=q,l \\ {\rm g}=1,2,3}} \Big( \overline{\Psi}{}^{\rm gf}_{\rm L}\,\gamma^\mu \check{D}_\mu \Psi^{\rm gf}_{\rm L} + \sum_{i=1,2} \overline{\Psi}{}^{\rm gf}_{i{\rm R}}\,\gamma^\mu \check{D}_\mu \Psi^{\rm gf}_{i{\rm R}} \Big) + \sum_{\substack{{\rm f}=q,l \\ {\rm g,g'}=1,2,3 \\ i=1,2}} \Big( {\bf Y}^{\rm f}_{\rm gg'}\,\overline{\Psi}{}^{\rm gf}_{\rm L}\Phi\Psi^{\rm g'f}_{i{\rm R}} + {\bf Y}'^{\rm f}_{\rm gg'}\,\overline{\Psi}{}^{\rm gf}_{\rm L}\widetilde{\Phi}\Psi^{\rm g'f}_{i{\rm R}} \Big) + {\rm h.c.} \\ &+ g^{\mu\nu}(\check{D}_\mu\Phi^\dagger)(\check{D}_\nu\Phi) + \tfrac{1}{2}(fM_{\rm P})^2(\check{S}_\mu)^2 - \lambda(\Phi^\dagger\Phi)^2 + \mu\, M_{\rm P}^2\,\Phi^\dagger\Phi - \xi M_{\rm P}^4 \bigg] \;, \end{aligned} \tag{2.6}$$

where the Hilbert action has been produced after we fix $\zeta = 1$, while $\check{D}_\mu$ implies the covariant derivatives in (2.5) with the $\varphi$ field set to zero. After all, the Weylon $\check{S}_\mu$ acquires the mass $fM_{\rm P}$, the compensator $\varphi$ is absorbed into the longitudinal component of $\check{S}_\mu$, and the potential terms are reduced to the Higgs potential in the SM $V(\Phi,\sigma) \longrightarrow \widehat{V}(\Phi) \equiv +\lambda(\Phi^\dagger\Phi)^2 - \mu M_{\rm P}^2\,\Phi^\dagger\Phi + \xi M_{\rm P}^4$, as in our previous papers [8][14][15].

We mention a subtlety about estimating the mass of $\check{S}_\mu$. Strictly speaking, the interpretation of the fourth term from the end in (2.6) as the mass term of $\check{S}_\mu$ is not quite correct. This is because, the longitudinal component of $S_\mu$ mixes not only with $\varphi$, but also with the Higgs field $\Phi$, after the $SU(2)$ breaking.

In order to clarify this more correctly, we perform a Weyl rescaling from the Jordan frame into Einstein frame with $\zeta$ fixed to be $\zeta = 1$ [16]:

$$\begin{aligned} e_\mu{}^m \;&\rightarrow \phi^{-1/2} e_\mu{}^m \;\;, \qquad g_{\mu\nu} \rightarrow \phi^{-1} g_{\mu\nu} \;\;, \qquad e \rightarrow \phi^{-2} e \;\;, \\ -\tfrac{1}{2} e\phi R \;&\rightarrow -\tfrac{1}{2}\kappa^{-2} eR + \tfrac{3}{4}\kappa^{-2}\phi^{-2} e g^{\mu\nu}(D_\mu\phi)(D_\nu\phi) + \kappa^{-2}\partial_\mu(eW^\mu) \;\;, \\ \phi \;&\equiv e^{-2\kappa\varphi}\left[1 + \beta\kappa^2 e^{2\kappa\varphi}(\Phi^\dagger\Phi)\right] \;\;. \end{aligned} \tag{2.7}$$

This rescaling gets rid of the multiplications of scalar terms in front of the scalar curvature. After this Weyl rescaling, the *bosonic terms* in the total action become

$$I_{\rm B} = \int d^4x\, e \bigg\{ -\tfrac{1}{2}\kappa^{-2} eR - \tfrac{1}{4}(U_{\mu\nu})^2 - \tfrac{1}{4}(W_{\mu\nu})^2$$

$$+\tfrac{1}{2}\widetilde{\phi}^{-1}(D_\mu\varphi)^2+3\left[(D_\mu\varphi)^2-\tfrac{1}{2}\kappa^{-1}\partial_\mu(\ln\widetilde{\phi})\right]^2$$
$$+\tfrac{1}{2}\left[D_\mu\widetilde{\Phi}^\dagger-\widetilde{\Phi}^\dagger\left\{\kappa D_\mu\varphi-\tfrac{1}{2}\partial_\mu(\ln\widetilde{\varphi})\right\}\right]^2$$
$$-\left[\lambda(\widetilde{\Phi}^\dagger\widetilde{\Phi})^2-\mu M_{\rm P}^2\widetilde{\phi}^{-1}(\widetilde{\Phi}^\dagger\widetilde{\Phi})+\xi M_{\rm P}^4\widetilde{\phi}^{-2}\right]\Bigg\}\ , \tag{2.8}$$

where $\widetilde{\Phi}$ and $\widetilde{\phi}$ are scale-invariant combinations defined by

$$\widetilde{\Phi}\equiv\phi^{-1/2}\Phi\ ,\qquad \widetilde{\phi}\equiv\left[1-\beta\kappa^2(\widetilde{\Phi}^\dagger\widetilde{\Phi})\right]^{-1}\ . \tag{2.9}$$

Note that when the Higgs field develop its v.e.v. $\widetilde{\Phi}_0\approx\mathcal{O}(M_{\rm H})$, the v.e.v. $\widetilde{\phi}_0$ of $\widetilde{\phi}$ will be of order $\mathcal{O}(1)$:

$$\widetilde{\phi}_0=\left[1-\beta\kappa^2(\widetilde{\Phi}_0^\dagger\widetilde{\Phi}_0)\right]^{-1}\approx\mathcal{O}(1)\ . \tag{2.10}$$

We now can estimate how the $\check{S}_\mu$ mass term is modified by the kinetic term of $\widetilde{\Phi}_0$. After expressing in terms of $\check{S}_\mu$, the $\widetilde{\Phi}$-kinetic term *does* have a contribution to the $\check{S}_\mu$ mass term, as $(\widetilde{\Phi}_0^\dagger\widetilde{\Phi}_0)\,\check{S}_\mu^2\approx M_{\rm H}^2\check{S}_\mu^2$, so that the modified (mass)$^2$ of $\check{S}_\mu$ is now

$$M_{\check{S}}^2=f^2M_{\rm P}^2\left[1+\frac{2}{f^2}\left(\frac{M_{\rm H}}{M_{\rm P}}\right)^2\right]\approx f^2M_{\rm P}^2\approx M_{\rm P}^2\ . \tag{2.11}$$

However, the modification compared with the first leading term of $\mathcal{O}(f^2M_{\rm P}^2)$ is negligible suppressed by the factor $(2/f^2)(M_{\rm H}/M_{\rm P})^2\approx10^{-32}$ for $f\approx\mathcal{O}(1)$, if fine-tuning of couplings is avoided due to arguments relating to naturalness.

Other good low-energy aspects in [8] are maintained here. For example, the right-handed neutrinos $\Psi_{1\rm R}^{1l}=\nu_{e\,\rm R}$, $\Psi_{1\rm R}^{2l}=\nu_{\mu\rm R}$, $\Psi_{1\rm R}^{3l}=\nu_{\tau\rm R}$ can be introduced into the SM for a seesaw mechanism. The relevant Yukawa couplings are [8]

$$L_\nu=\sum_{\substack{\rm g,g'=1,2,3\\ i=1}}\left(\mathbf{Y}^l_{\rm gg'}\overline{\Psi}{}^{{\rm g}l}_{\rm L}\Phi\Psi^{{\rm g}'l}_{i\rm R}+{\rm h.c.}+\tfrac{1}{2}\mathbf{Y}^{\rm RR}_{\rm gg'}\Psi^{{\rm g}l}_{i\rm R}{}^TC\sigma\Psi^{{\rm g}'l}_{i\rm R}\right)\ . \tag{2.12}$$

In the frame $\sigma=M_{\rm P}$, *i.e.*, $\Delta=\sqrt{2}M_{\rm P}$, there are super-heavy Majorana masses for the right-handed neutrinos. The subsequent $SU(2)$ breaking gives Dirac masses connecting the left- and right-handed neutrinos with the familiar $6\times6$ mass matrix

$$\mathbf{M}_\nu=\frac{1}{\sqrt{2}}\begin{pmatrix}\mathbf{0} & \eta\mathbf{Y}^l_{\rm gg'}\\ \eta\mathbf{Y}^l_{\rm g'g} & \sqrt{2}M_{\rm P}\mathbf{Y}^{RR}_{\rm gg'}\end{pmatrix}\ , \tag{2.13}$$

where $\eta \equiv \sqrt{2\mu {M_{\rm P}}^2/\lambda} \simeq \mathcal{O}(250\,{\rm MeV})$ is the $SU(2)$ breaking scale. Six seesaw masses come out as the eigenvalues of this matrix, yielding the three light neutrinos and three heavy neutrinos. The scale of right-handed neutrino masses is directly related to $M_{\rm P}$. The absence of right-handed light neutrino is thus attributed to the super-heavy mass of $\mathcal{O}(M_{\rm P})$ [8].

## 3. Chaotic Inflation

Our model has an additional good feature of accommodating chaotic inflation with no extra field. Chaotic inflation *via* Higgs doublet in the SM has been recently discussed in [17]. However, our model is distinguished from the latter due to *local* scale invariance in the system.

Even though our original model in [8] does not address inflation, it contains all the ingredients necessary to accommodate the chaotic inflationary scenario. The standard model Higgs particle $h$ serves as the inflaton field with a strong non-minimal coupling $\beta$ to gravity. Our basic action is treated in the physical gauge with the following interaction terms [15]:

$$I = \int d^4x\, e \left[\widehat{\mathcal{L}} - \tfrac{1}{2}(\beta h^2 + M_{\rm P}^2)R + \tfrac{1}{2}g^{\mu\nu}(\partial_\mu h)(\partial_\nu h) - \tfrac{1}{4}\lambda(h^2-\eta^2)^2 - M_{\rm P}^4\Big(\xi - \tfrac{1}{4}\mu^2\lambda^{-1}\Big)\right] . \quad (3.1)$$

Here we adopt the particular gauge $\varphi = 0$ as before, and $\widehat{\mathcal{L}}$ contains the SM and the massive Weylon $(\breve{S}_\mu)$ particle interactions. In the physical gauge the effective real Higgs field $h \equiv \sqrt{2}\,\mathcal{R}e\,\Phi^0$ serves as the *inflaton* developing a non-zero v.e.v. After inflation $h$ settles in the minimum of its potential at the symmetry breaking scale $\eta = \sqrt{2\mu M_{\rm P}^2/\lambda}$. Compared with $M_{\rm P}^2/\beta$, $\eta$ is small, but large enough to render $h$ massive. Present day gravitational interactions are mediated by the effective Planck mass squared $M_{\rm eff}^2 = M_{\rm P}^2 + \beta\eta^2$ such that $M_{\rm eff} \simeq M_{\rm P}$ is maintained to a very good approximation.

Desirable inflationary scenario requires $\beta \gg 1$. We can perform a new Weyl rescaling $g_{\mu\nu} \;\to\; (1+\beta\kappa^2h^2)\,g_{\mu\nu}$ in order to reach the physical frame in which the $h^2$ interactions with the scalar curvature are absent. This is a Weyl rescaling for field redefinitions that is separate from our original local scale transformation $\widetilde{U}(1)$, because the latter is now 'fixed' under our particular gauge $\varphi = 0$. The action $I$ in the physical frame is

$$I \longrightarrow \int d^4x\; e\;\left[\widetilde{\mathcal{L}} - \tfrac{1}{2}M_{\rm P}^2 R + \tfrac{1}{2}g^{\mu\nu}(\partial_\mu \mathcal{H})(\partial_\nu \mathcal{H}) - \tfrac{1}{4}\lambda(1+\beta\kappa^2h^2)^{-2}(h^2-\eta^2)^2\right] \;, \quad (3.2)$$

where $g_{\mu\nu}$, $R$ and $\mathcal{H}$ are all calculated in the new frame. The $\mathcal{H}$ is given in terms of $h$ as $d\mathcal{H}/dh = \sqrt{1+\beta\kappa^2h^2+6\beta^2\kappa^2h^2}/(1+\beta\kappa^2h^2)$, whose *exact* solution is [15]

$$\mathcal{H} = M_{\rm P}\sqrt{\frac{6\beta+1}{\beta}}\,\cosh^{-1}\sqrt{\beta\kappa^2(6\beta+1)h^2+1} - \frac{\sqrt{6}M_{\rm P}}{2}\,\ln\left[\frac{\sqrt{\beta(6\beta+1)h^2+M_{\rm P}^2}+\sqrt{6}\beta h}{\sqrt{\beta(6\beta+1)h^2+M_{\rm P}^2}-\sqrt{6}\beta h}\right] . \quad (3.3)$$

In the paradigm of inflationary scenario, initially the inflaton field is larger than $M_{\rm P}$ and slow rolls down the potential, signifying the inflationary phase characterized by the rapid exponential expansion of the universe. The end of inflation occurs when the inflaton reaches the minimum of the potential where it loses energy via rapid oscillations. The energy released results in particle production that interact strongly and come to thermal equilibrium at some temperature $T^\star$ also known as the reheat temperature. The latter is restricted to be $\ll 2.8\times10^{16}$ GeV [18] to respect the WMAP bound on tensor fluctuations [19]. The two regions of interest are when $h \ll M_{\rm P}/\sqrt{\beta}$ and $h \gg M_{\rm P}/\sqrt{\beta}$. Our solution for $\mathcal{H}$ implies that in the first case $\mathcal{H}\simeq h$, while in the second case $h\simeq (M_{\rm P}/\sqrt{\beta})\,\exp\left[\mathcal{H}/(\sqrt{6}M_{\rm P})\right]$. In these two extreme cases the potential takes the following forms

$$V(h)\simeq\begin{cases}\dfrac{\lambda M_{\rm P}^4}{4\beta^2}\left(1-\dfrac{2M_{\rm P}^2}{\beta h^2}\right)^{-2}\longrightarrow\dfrac{\lambda M_{\rm P}^4}{4\beta^2} & (\text{for } h\gg M_{\rm P}/\sqrt{\beta})\ ,\\ \dfrac{1}{4}\lambda h^4 & (\text{for } \eta\ll h< M_{\rm P}/\sqrt{\beta})\ .\end{cases} \quad (3.4)$$

The customary slow roll parameters $\widehat{\varepsilon}$, $\widehat{\eta}$ and $\widehat{\delta}$ [20] in our model are [15]

$$\widehat{\varepsilon}=\frac{M_{\rm P}^2}{2V^2}\left(\frac{dV}{d\mathcal{H}}\right)^2\simeq\frac{4M_{\rm P}^4}{3\beta^2h^4}\ ,\quad \widehat{\eta}=\frac{M_{\rm P}^2}{V}\left(\frac{d^2V}{d\mathcal{H}^2}\right)\simeq-\frac{4M_{\rm P}^2}{3\beta h^2}\ ,$$
$$\widehat{\delta}^2=\frac{M_{\rm P}^4}{V^2}\left(\frac{d^3V}{d\mathcal{H}^3}\right)\left(\frac{dV}{d\mathcal{H}}\right)\simeq\frac{16M_{\rm P}^4}{9\beta^2h^4}\ . \quad (3.5)$$

Slow roll ends when $\widehat{\epsilon}\simeq1$, so $h_{\rm end}\simeq(4/3)^{1/4}M_{\rm P}/\sqrt{\beta}$ at the end of inflation. The e-foldings number in the inflation era, when $h$ evolves from $h$ to $h_{\rm end}$, is

$$N=-\int_h^{h_{\rm end}}\frac{1}{M_{\rm P}}\frac{1}{\sqrt{2\widehat{\epsilon}}}d\mathcal{H}\simeq\frac{3\beta}{4M_{\rm P}^2}(h^2-h_{\rm end}^2)\ . \quad (3.6)$$

The numerical value of e-foldings required depends on the COBE normalization [21]. With $N\equiv N_{\rm COBE}=60$ and $h\equiv h_{\rm COBE}=4N_{\rm COBE}M_{\rm P}/(3\sqrt{\beta})$, we get $h_{\rm COBE}^2/h_{\rm end}^2\simeq 4N_{\rm COBE}/3\gg1$. The spectral index $n_{\rm s}$, the ratio $r$ of the tensor to scalar perturbations, and the spectral index running $n_{\rm r}$ can now be calculated from $n_{\rm s}=1-6\widehat{\epsilon}+2\widehat{\eta}$ , $r=16\widehat{\epsilon}$ , $n_{\rm r}=16\widehat{\epsilon}\widehat{\eta}-24\widehat{\epsilon}^2-2\widehat{\delta}^2$ for $h\simeq h_{\rm COBE}$ and at wave number $k\simeq0.002{\rm Mpc}^{-1}$ [22].

We find $n_{\rm s} \simeq 1 - 8(4N_{\rm COBE} + 9)/(4N_{\rm COBE})^2 \simeq 0.97$ and $r \simeq 12/(N_{\rm COBE})^2 \simeq 0.0033$ and $n_{\rm r} \simeq -2/N_{\rm COBE} = -0.0006$ and fall in the acceptable regime of parameter space.

## 4. Concluding Remarks

Our model has, in a sense, solved the long-standing puzzle about the breaking of local scale invariance at the *classical* level without Higgs mechanism. It has been well known that scale invariance [23] or conformal invariance [24] can be broken by quantum corrections. What we have shown above is that by the Stückelberg mechanism [2], Weyl's local scale invariance [1] is broken at the *classical* level. In particular, this breaking is *neither* explicit *nor* artificially put 'by hand' [8].

The main aspects in our original papers [8] are intact, such as the Weylon not coupling to fermions in their kinetic terms. This also implies the absence of $\widetilde{U}(1)$ anomaly. Even though our potential terms are reduced exactly to the SM Higgs potential, the Weylon still couples to the Higgs doublet $\Phi$. However, the Weylon- $\Phi$ couplings are different from vector-fermion minimal couplings, because of either derivative couplings *e.g.*, $f\Phi^\dagger S^\mu \partial_\mu \Phi$ or the two-Weylon coupling $f^2 \Phi^\dagger \Phi S^\mu S_\mu$. Hence its effective coupling is suppressed by (momentum)$/M_{\rm P}$, and is very hard to be detected by the near-future collider experiments.[7)]

In our system, the 'dilaton' $\varphi$ automatically becomes a compensator. No matter how many general complex scalars are present, it is always one real scalar singled out that becomes the compensator. Hence in any locally scale-invariant system with a gauge field $S_\mu$, if at least one real scalar is present, that scalar becomes a compensator, and local invariance is *necessarily* broken by the Stückelberg mechanism [2].

The identification of a 'dilaton' with a compensator is also motivated by our recent success of axion and 'dilaton' regarded as compensators with $N = 1$ supersymmetry [25][26]. The success with both global [25] and local [26] supersymmetry provides supporting evidence for the consistency of identifying the 'dilaton' with a Stückelberg compensator [2].

Lastly we address the issue of quantum corrections in our model. There are several issues to consider. First, gravitational interactions are non-renormalizable. Second, one's

[7)] Notice, however, that the mass factor in (2.1) can be arbitrary, but *not* necessarily $M_{\rm P}$, even though the latter is the most natural scale. That mass can be much lighter than $M_{\rm P}$, and so can be the Weylon mass more detectable in the near future. The Hilbert action is still induced by an appropriate value of $\zeta$.

prejudice in the choice of the physical frame as opposed to the choice of any other frame is ambiguous. Third, the choice of a particular gauge in calculating physically relevant quantities may induce ambiguities. Fourth, there remains the problem of choosing the cutoff. In the absence of a clear procedure, we follow the philosophy advocated by Donoghue [27] and further elaborated upon by Robinson and Wilczek [28]. We treat our model as an effective theory of gravity. In this regard, quantum corrections are computed in the Einstein frame. One loop quantum corrections due to gravitational and other SM interactions give small contributions relative to the physical quantities calculated at the tree level. Our main results will only be marginally shifted due to these corrections, but the broader features of the model will still be upheld.

Fifth, we need to mention the issue of trace (conformal) anomaly, not arising from the chiral fermion loops, that is different from the one mentioned above. Trace anomaly can arise from various fields with various spins. Even though we do not address ourselves to this issue in this paper, we cite the works [29][30][12] for basic trace anomaly computations, or the paper by Christensen and Duff [31], where various trace anomaly coefficients are listed. Since some fields, such as the third-rank field $\phi_{\mu\nu\rho}$, have anomaly coefficients with signs opposite to those of the graviton $e_\mu{}^m$ and the quarks and leptons, we have good chance that the total trace anomaly is cancelled.